%% file: main.tex
\documentclass[sigconf]{acmart}

\usepackage[ruled,vlined]{algorithm2e}
\usepackage{adjustbox}
\usepackage{comment}
\usepackage[normalem]{ulem}
\usepackage{xcolor}
\usepackage{makecell}
\usepackage{multirow}
\usepackage{enumitem}
\usepackage[table]{xcolor}
\usepackage{arydshln}
\usepackage{float} 
\usepackage{placeins}
\usepackage{booktabs}
\usepackage{tabularx}
\usepackage{enumitem}
\usepackage{subcaption} 
\usepackage{caption}
\SetCommentSty{mycommentstyle}

\SetCommentSty{cameraready}

\newcommand{\algname}{CREAM}

\AtBeginDocument{%
  }

\makeatletter
\def\@specialsection#1{%
  \let\@vspace\@vspace@orig
  \let\@vspacer\@vspacer@orig
  \ifcase\ACM@format@nr
    \par\medskip\noindent#1: %
  \or
    \par\medskip\noindent#1: %
  \or
    \par\medskip\noindent#1: %
  \or
    \par\medskip\noindent#1: %
  \or 
    \par\smallskip\noindent\textbf{#1}\par\nobreak\smallskip
  \fi
  \let\@vspace\@vspace@acm
  \let\@vspacer\@vspacer@acm
}
\makeatother

\setcopyright{acmlicensed}
\copyrightyear{2026}
\acmYear{2026}
\setcopyright{cc}
\setcctype{by}
\acmConference[KDD '26]{Proceedings of the 32nd ACM SIGKDD Conference on Knowledge Discovery and Data Mining V.1}{August 09--13, 2026}{Jeju Island, Republic of Korea}
\acmBooktitle{Proceedings of the 32nd ACM SIGKDD Conference on Knowledge Discovery and Data Mining V.1 (KDD '26), August 09--13, 2026, Jeju Island, Republic of Korea}
\acmPrice{}
\acmDOI{10.1145/3770854.3780281}
\acmISBN{979-8-4007-2258-5/2026/08}

\begin{document}

\title{CREAM: Continual Retrieval on Dynamic Streaming Corpora with Adaptive Soft Memory}

\settopmatter{authorsperrow=4}
\author{HuiJeong Son}
\email{huijeong.son@korea.ac.kr}
\affiliation{
  \institution{Korea University}
  \country{Seoul, Korea}
}
\author{Hyeongu Kang}
\email{hyeongu_kang@korea.ac.kr}
\affiliation{
  \institution{Korea University}
  \country{Seoul, Korea}
}
\author{Sunho Kim}
\email{sunho_kim@korea.ac.kr}
\affiliation{
  \institution{Korea University}
  \country{Seoul, Korea}
}
\author{Subeen Ho}
\email{hosubin02@korea.ac.kr}
\affiliation{
  \institution{Korea University}
  \country{Seoul, Korea}
}
\author{SeongKu Kang}
\email{seongkukang@korea.ac.kr}
\affiliation{
  \institution{Korea University}
  \country{Seoul, Korea}
}
\author{Dongha Lee}
\email{donalee@yonsei.ac.kr}
\affiliation{
  \institution{Yonsei University}
  \country{Seoul, Korea}
}

\author{Susik Yoon}
\email{susik@korea.ac.kr}
\affiliation{%
  \institution{Korea University}
  \country{Seoul, Korea}
}

\renewcommand{\shortauthors}{HuiJeong Son et al.}

\begin{abstract}
Information retrieval (IR) in dynamic data streams is a crucial task, as shifts in data distribution degrade the performance of AI-powered IR systems. To mitigate this issue, memory-based continual learning has been widely adopted for IR. However, existing methods rely on a fixed set of queries with ground-truth documents, which limits generalization to unseen data, making them impractical for real-world applications. To enable more effective learning with unseen topics of a new corpus without ground-truth labels, we propose CREAM, a self-supervised framework for memory-based continual retrieval. CREAM captures the evolving semantics of streaming queries and documents into dynamically structured soft memory and leverages it to adapt to both seen and unseen topics in an unsupervised setting. We realize this through three key techniques: fine-grained similarity estimation, regularized cluster prototyping, and stratified coreset sampling. 
Experiments on two benchmark datasets demonstrate that CREAM exhibits superior adaptability and retrieval accuracy, outperforming the strongest method in a label-free setting by 27.79\% in Success@5 and 44.5\% in Recall@10 on average, and achieving performance comparable to or even exceeding that of supervised methods.

\end{abstract}



%


\maketitle
\vspace*{-6pt}
\newcommand\kddavailabilityurl{10.6084/m9.figshare.30957539}
\ifdefempty{\kddavailabilityurl}{}{
\begingroup\small\noindent\raggedright\textbf{Resource Availability:}\\
The source code of this paper has been made publicly available at \url{\kddavailabilityurl}.
\endgroup
}

\input{tex/01-Introduction}
\input{tex/02-RelatedWork}
\input{tex/03-ProblemSetting}
\input{tex/04-Methodology}

\input{tex/05-Experiments}
\input{tex/06-Conclusion}

\section*{Acknowledgments}
This work was partly supported by the Institute of Information \& Communications Technology Planning \& Evaluation (IITP)-ICT Creative Consilience Program (IITP-2026-RS-2020-II201819), IITP-ITRC (Information Technology Research Center) (IITP-2026-RS-2024-00436857), Artificial Intelligence Star Fellowship Program (IITP-2026-RS-2025-02304828), and the National Research Foundation of Korea (NRF) (RS-2024-00406320).

\balance
\bibliographystyle{unsrt}
\bibliography{References}

\input{tex/07-Appendix}
\end{document}

%% file: tex/01-Introduction.tex
\begin{figure}[t]
\includegraphics[width=\columnwidth]{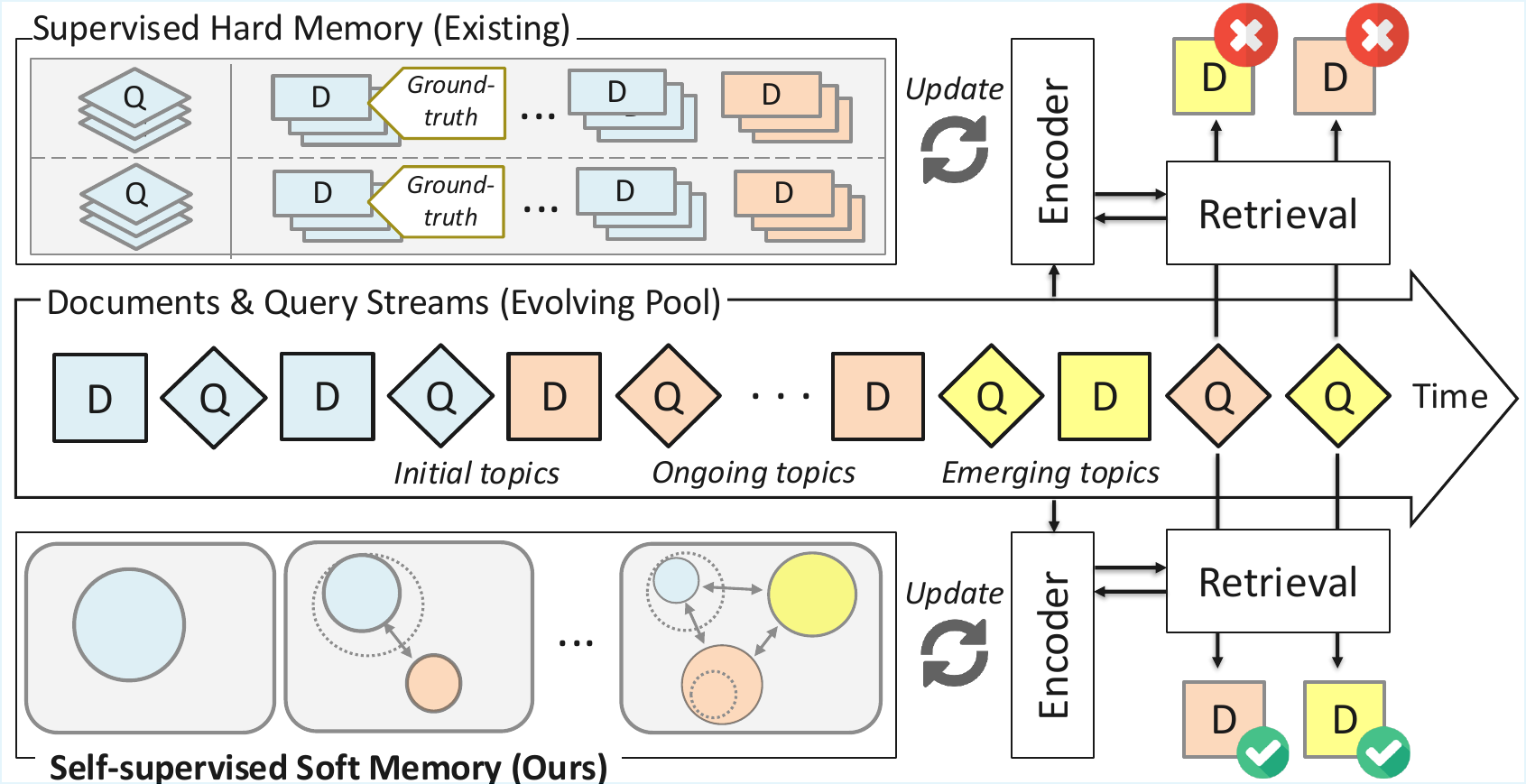}
    \vspace{-0.5cm}
    \caption{Comparison of existing (top) and our (bottom) approaches for memory-based continual retrieval.}
    \vspace{-0.5cm}
    \label{fig:motivation}
\end{figure}

\section{Introduction}
\subsection{Background}
Information retrieval (IR) in online environments, powering real-time retrieval-augmented generation~\cite{lewis2020retrieval} and agentic context engineering~\cite{zhang2025agentic}, is emerging as a key technology for various downstream applications. For example, in a real-time news summarization system~\cite{pdsum}, a query “\textit{What are the current issues in the global supply chain?}” would require retrieving relevant articles covering current events, such as geopolitical conflicts or new tariff policies, at the time of the query. Identifying relevant documents with dynamically evolving topics is challenging, as pretrained retrieval models become outdated under domain shifts. This challenge is especially critical in real-world IR systems requiring timely and accurate responses, which is more pronounced in emerging agentic AI frameworks that facilitate real-time decision-making~\cite{agentic_ai}. 


Specifically, consider a scenario where topics gradually shift from the medical domain to the business domain. In a query “\textit{Has the agent been approved?}”, the term \textit{agent} typically refers to a drug in the medical domain, whereas it denotes a person or agency in the business context. If an IR system has not adapted to the new domain, it may return irrelevant medical documents, potentially leading to an incorrect answer such as “\textit{The FDA approved the therapeutic agent.}” In contrast, if a system rapidly adapts to the emerging domain while retaining relevant knowledge from the previous domain, it is capable of retrieving appropriate business-related documents and producing a more appropriate answer, such as “\textit{The licensing board approved the real estate agent’s application.}” 


\subsection{Existing Efforts}
In a typical IR system, an encoder is optimized to enhance the semantic similarity between query-document pairs labeled as relevant, with these labels obtained through human annotation. As illustrated in Figure \ref{fig:motivation}, when the distribution of queries and documents evolves over time with diverse topics, matching relevant pairs becomes increasingly difficult unless the encoder is continually updated to reflect the evolving corpora. In practice, web-scale corpora in typical IR systems involve a continuous influx of documents and queries, making retraining on all past data for training computationally inefficient and often infeasible. The naive incremental update of the encoder, however, suffers from catastrophic forgetting, a well-known issue in deep learning where previously acquired knowledge is overwritten by new information~\cite{catastrophic_forgetting}. To address this challenge, existing continual retrieval methods adopt memory-based continual learning strategies~\cite{er, mir, gss, ocs, l2r, clever} to acquire new knowledge without forgetting the old one. 

As shown in the upper part of Figure~\ref{fig:motivation}, existing methods with memory-based continual learning strategies employ dedicated storage for a fixed set of given queries and their corresponding ground-truth documents, which can be referred to as \emph{hard memory}. Query-document pairs are sampled from hard memory to update the encoder, while new documents relevant to the predefined queries are added from the streaming corpora. This approach has proven effective in scenarios where the topical distribution of the corpus remains relatively stable and consistent with the predefined query set~\cite{l2r}. However, simply reusing shift-unaware, predefined query-document pairs stored in hard memory for continual training can cause the encoder to learn information that is less relevant to the current topic distribution, leading to poor adaptation to distributional changes. Moreover, in real-time applications~\cite{ustory}, human-curated supervision from a set of fixed queries with ground-truth documents is not always available in a timely manner~\cite{human_ai}, making the hard memory strategy impractical. As a result, such methods fail to support effective retrieval on newly emerging topics, and may degrade performance on the initial or ongoing topics.

\subsection{Main Idea and Contributions}
To address these practical limitations, we propose a novel concept called \emph{soft memory} that can adapt to the ever-changing topical distributions of queries and documents. Soft memory dynamically tracks relevant queries and documents across varying topics, making it better than hard memory for adapting to evolving topic distributions, especially without supervision.
As illustrated in the bottom part of Figure~\ref{fig:motivation}, semantically similar queries and documents in the streaming corpora are continuously grouped and expanded within the memory. For example, the soft memory may begin with creating a group of documents and queries representing initial topics, then add groups related to ongoing topics, and eventually incorporate new groups for emerging topics while phasing out older ones. By leveraging its dynamic structural representation of evolving topic distributions, soft memory enables self-supervised training \textit{without relying on predefined queries or ground-truth relevant documents}, providing high-quality pseudo-labeled samples to update the encoder in line with topic shifts. Ultimately, this results in more accurate retrieval aligned with the latest topics.

To instantiate the soft memory strategy for a more practical and effective continual retrieval system, we propose \textbf{\algname{}}, a framework for \underline{C}ontinual \underline{RE}trieval with \underline{A}daptive Soft \underline{M}emory. While the soft memory is fundamentally adequate for memory-based continual learning to address the dynamic distributional shift in queries and documents, there remain non-trivial challenges in integrating this concept into the reliable continual retrieval pipeline. To this end, \algname{} is built upon the following three core techniques:

\begin{itemize}[leftmargin=9pt, noitemsep]
\item{\textbf{Fine-grained similarity estimation:}} In the absence of external supervision from labeled query-document pairs, a simple similarity estimation based on conventional single-vector representations is insufficient to capture the complex and evolving semantics of corpora. This limitation is practically significant, as self-supervision with noisy relevance signals can lead to critical degradation of the encoder. Thus, we fully exploit the entire token-level information to compute fine-grained semantic similarities both for memory construction and retrieval, inspired by the contextualized late interaction~\cite{colbertv2}.  This enables the encoder to robustly adapt to subtle contextual shifts and emerging topics, even without direct supervision signals. 
\item{\textbf{Regularized cluster prototyping:}} We perform streaming clustering of queries and documents with high fine-grained similarity to structure a soft memory. However, variations in token lengths of corpora incur significant overhead in token-level similarity computations, in addition to the cost of pairwise comparisons for cluster assignment. To achieve efficient yet accurate incremental clustering, we represent each cluster using a prototype (i.e., a centroid) regularized in a fixed token length. Specifically, we leverage locality-sensitive hashing to normalize the embedding sizes of the corpora, enabling semantically fine-grained prototypes while preserving alignment with theoretically bounded information loss.

\item{\textbf{Stratified coreset sampling:}} The soft memory serves as an effective pool of pseudo-labeled query-document pairs. We aim to select a diverse set of representative query-document training samples to ensure the encoder reflects a comprehensive knowledge space in the soft memory. To this end, we employ stratified sampling to construct a coreset of samples that efficiently and effectively preserves the semantic diversity of the soft memory. This coreset is used to train the encoder in a self-supervised manner with the contrastive objective, promoting generalization of varying query and document semantics. 
\end{itemize}

\sloppy
In summary, our main contributions are as follows:
\begin{itemize}[leftmargin=10pt, noitemsep]
    \item We propose a novel concept of soft memory for memory-based continual learning in IR systems, aimed at practically addressing unbounded, unlabeled, and topic-shifting streaming corpora.
    \item We present \algname{}, the first continual retrieval framework that operates in a fully unsupervised setting, incorporating three key technical ingredients, fine-grained similarity, regularized cluster prototypes, and stratified coreset samples, that collectively facilitate robust self-supervision of the encoder throughout continual learning. The source code is publicly available at \url{https://github.com/DAIS-KU/CREAM}.  
    \item On two extensive real-world datasets, \algname{} achieves superior retrieval performance, surpassing the strongest baseline by 27.79\% in Success@5 and 44.5\% in Recall@10 on average.
\end{itemize}

%% file: tex/02-RelatedWork.tex
\section{Related Work}
\subsection{Information Retrieval}
Traditional information retrieval approaches are often categorized into sparse, dense, and generative retrieval. Sparse retrieval (SR), such as BM25~\cite{bm25}, relies on term frequency and inverse document frequency to compute relevance scores based on exact token matches. While efficient and interpretable, these methods suffer from key limitations: they rely heavily on exact string matches and fail to capture the contextual nuance of semantically similar expressions. As a result, they often underperform in settings where lexical variation or richer semantic understanding is required. 

Dense retrieval (DR) addresses these issues by encoding queries and documents into dense vector representations using an encoder. 
Cross-encoders~\cite{bert_reranker} jointly encode the concatenated query and document, employing a final linear layer to map the aggregate sequence representation to a scalar similarity score. Although highly effective, this approach requires pairwise computation between all query-document pairs, which is computationally expensive. 
In contrast, dual-encoders~\cite{dpr} independently encode queries and documents, enabling fast retrieval via cosine similarity on precomputed embeddings. While this method significantly reduces inference time, it may struggle with fine-grained matching due to the lack of deep interaction between the query and the document. To address this trade-off, ColBERT~\cite{khattab2020colbert} has introduced late interaction, which balances efficiency and efficacy by delaying fine-grained interactions until the retrieval stage. 

With the rise of generative models, generative retrieval (GR) has emerged as a new concept in IR. Differentiable Search Index (DSI)~\cite{dsi} first proposed the concept of GR which the model generates document identifiers auto-regressively given a query. They showed that larger models achieved greater performance gains, but also that the gains diminished on larger corpora. While DR and GR leverage language models and share the high-level goal of retrieving relevant documents given a query, they formulate the retrieval problem differently. DR focuses on accurately computing and ranking the similarity between queries and documents, whereas GR aims to generate the correct document identifier for a given query, making direct comparison between them inadequate.

\subsection{Continual Learning on IR}
In a continual retrieval setting, where new documents and queries continuously arrive, a retrieval model needs to be repeatedly updated to return the most relevant documents for a given query in the latest context. Naively updating the model with the new data can lead to catastrophic forgetting, particularly for DR and GR models, where the models overwrite previously acquired knowledge and lose their generalizability on earlier data. Among the various continual learning strategies that can be applied to mitigate this issue, Memory replay~\cite{er, mir, gss, ocs} has been widely adopted due to its simplicity and effectiveness, especially when combined with a contrastive learning objective~\cite{scstory}. These memory-based approaches typically maintain an external memory initialized with a set of queries and their corresponding relevant document pairs (i.e., positive samples). As new documents arrive, the encoder is updated to make each query closer to the positive samples while pushing it away from irrelevant documents (i.e., negative samples). Recent works such as L2R~\cite{l2r} and CLEVER~\cite{clever} represent state-of-the-art frameworks for DR- and GR-based continual retrieval, respectively. L2R defines negativity and diversity metrics to mitigate false negatives during negative sampling. CLEVER leverages external memory for a pseudo-query generator with a reconstruction-based objective, a regularization-based objective, and dynamic indexing via Incremental Product Quantization (IPQ). However, both approaches present practical limitations in their supervision assumption: L2R requires a fixed set of queries along with ground-truth documents for training, and CLEVER also depends on a large number of positive query-document pairs for effective model initialization.


%% file: tex/03-ProblemSetting.tex
\section{Problem Setting}
\label{sec:problem_setting}
Let $S_t = (Q_t, D_t)$ be a query-document stream during a session $t$, where a set $Q_t$ of queries and a set $D_t$ of documents are associated with diverse domains or topics, the compositions of which evolve over time. Then, formally, the retrieval task $\mathcal{R}$ at session $t$ to find the set of relevant documents $D^{rel}_t \subset D_t$ for the given queries $Q_t$ using an encoder $f_{t-1}$ is formulated as:
\begin{equation}
\label{eq:task_definition}
\mathcal{R}: \langle f_{t-1}, Q_t, D_t \rangle \rightarrow D^{rel}_t.
\end{equation}

To adapt to the evolving distribution of data over time, a memory-based continual learning algorithm $\mathcal{A}$ updates the encoder with a memory $M$. At each session $t$, the algorithm $\mathcal{A}$ selects training samples from both the current stream $S_t$ and the previous memory $M_{t-1}$ to update the encoder $f_{t-1}$. It then produces an updated encoder and memory:
\begin{equation}
\label{eq:problem_setting}
\mathcal{A} : \langle f_{t-1}, M_{t-1}, S_t \rangle \rightarrow \langle f_t, M_t \rangle.
\end{equation}


When evaluating the retrieval performance under the continual learning algorithm $\mathcal{A}$, two evaluation protocols can be considered, depending on the composition of the training and test sets. The disjoint setting follows a standard machine learning protocol that separates the training samples from the test samples. The shared setting adopts an IR-specific protocol where the same document corpus applies to both the training and testing phases. Formally, the disjoint evaluation $\mathcal{R}^*_{disjoint}$ separates test queries $Q^*_t$ and documents $D^*_t$ from the training queries $Q_t$ and documents $D_t$, whereas the shared-pool evaluation $\mathcal{R}^*_{shared}$ uses the same document pool for the disjoint train and test queries:
\begin{equation}
\label{eq:trditional_evaluation}
\mathcal{R}^{*}_\text{disjoint}: \langle f_t, Q^{*}_t, D^{*}_t \rangle \rightarrow D^{rel*}_t.
\end{equation} 
\begin{equation}
\label{eq:staticir_evaluation}
\mathcal{R}^{*}_\text{shared}: \langle f_t, Q^{*}_t, D_t \rangle \rightarrow D^{rel*}_t.
\end{equation}


%% file: tex/04-Methodology.tex
\begin{figure*}[t]
    \centering
    \includegraphics[width=\textwidth]{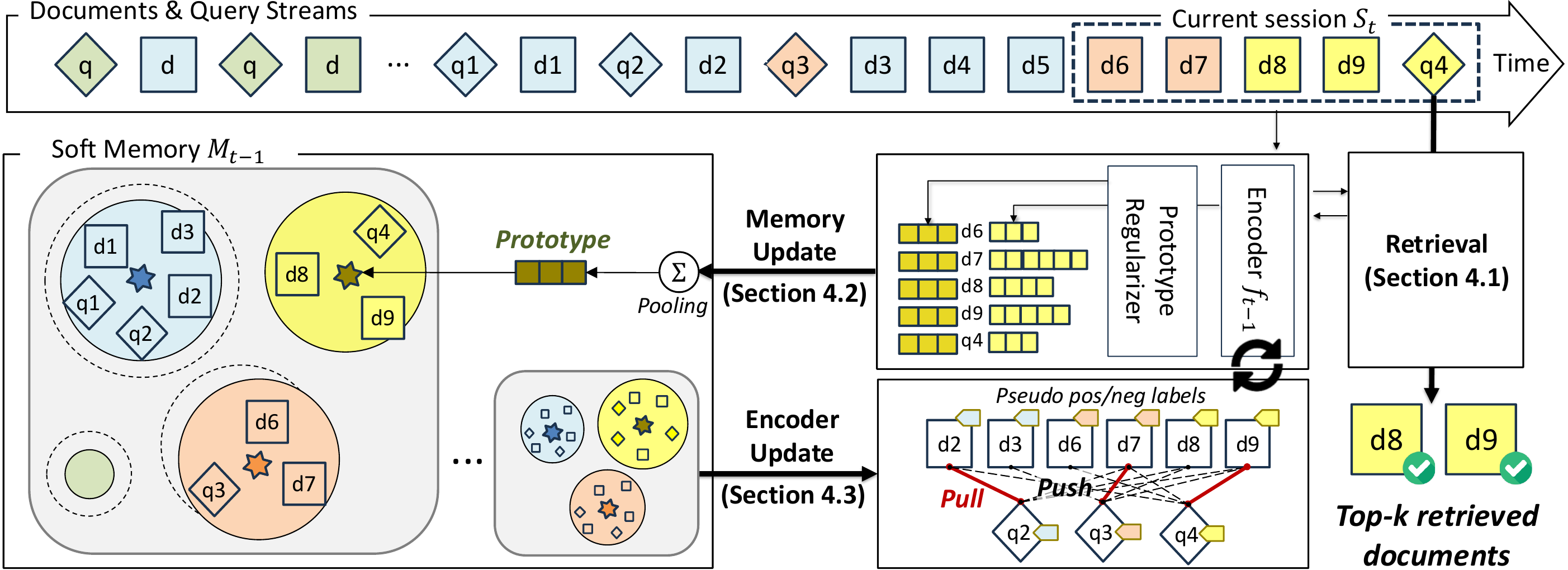}
    \caption{Overall framework of \algname{} with three components: (1) a retrieval component that returns the relevant documents to a given query with the up-to-date encoder; (2) a memory update component that captures the recent knowledge while preserving previously acquired information through streaming clustering with regularized prototypes; and (3) an encoder update component that facilitates self-supervised training using contrastive objective and the structure of soft memory.}
    \label{fig:overview}
\end{figure*}

\begin{algorithm}[t]
\LinesNumbered  
\caption{Overall Procedures of \algname{} (Section \ref{sec:method_overview})}
\label{overview}
\KwIn{Documents $D_t$, Queries $Q$, Encoder $f_t$, Memory $M_t$, Sessions $t \in T$}
\KwOut{Updated Encoder $f_t$, Updated Memory $M_t$, Retrieved Result $D^{rel}_t$}
\For{each session $t \in \{1, \dots, T\}$}{
    $f_t \gets f_{t-1}$, $M_t \gets M_{t-1}$, $D^{rel}_t \gets \emptyset$\\
    \tcc{\texttt{Retrieval (Section \ref{sec:retrieval})}}
    $D^{rel}_t \gets \text{Retrieve}(f_t, Q_t, M_t)$\\
    \tcc{\texttt{Memory Update (Section \ref{sec:memory_update})}}
    $M_t \gets \text{AssignToCluster}(f_t, Q_t, D_t, M_t)$\\
    $M_t \gets \text{UpdateClusterSummary}(f_t, M_t)$ \\
    \tcc{\texttt{Encoder Update (Section \ref{sec:encoder_update})}}   
    $S_q \gets \text{SelectTrainingQueries}(f_t, M_t)$\\
    $S_p, S_n \gets \text{SelectTrainingSamples}(f_t, S_q, M_t)$\\
    $f_t \gets \text{UpdateEncoder}(f_t, S_q, S_p, S_n)$\\
}
\KwRet{$f_t, D^{rel}_t$}
\end{algorithm}

\section{Methodology}
\label{sec:method_overview}
Algorithm~\ref{overview} and Figure~\ref{fig:overview} outline the proposed framework \algname{}, supporting three main operations: (i) the Retrieval stage, which returns relevant documents for incoming queries (Line 3); (ii) the Memory Update stage, which updates the existing memory by incrementally grouping the new queries and documents in clusters through streaming clustering (Lines 4 and 5); and (iii) the Encoder Update stage, which selects training documents per query from the latest memory to update the encoder (Lines 6-8). The following sections describe each step in detail.

\subsection{Retrieval}
\label{sec:retrieval}
    An effective retrieval system requires capturing the subtle contextual semantics between queries and documents, which is more pronounced in a label-free setting for continual retrieval. To minimize the loss of information from token embeddings of queries and documents, we aim to preserve token-level granularity in relevance estimation. Specifically, instead of relying on a single embedding derived from mean pooling or the [CLS] token, we adopt a token-level similarity approach inspired by ColBERT’s late interaction~\cite{colbertv2} to preserve the semantic granularity of individual tokens. Unlike ColBERT, however, we do not modify the encoder architecture or use special tokens to maintain simplicity and efficiency.

Given token embedding sequences $E_q \in \mathbb{R}^{n \times l}$ of a query $q$ and $E_d \in \mathbb{R}^{m \times l}$ of a document $d$, where $n$ and $m$ are the number of tokens and $l$ is the embedding dimension, we take the sum of the maximum cosine similarities of each token of a query and all tokens in document to get the token-level similarity $Sim_{qd}$:
\begin{equation}
\label{similarity}
   Sim_{qd}= \sum_{i \in [\![E_q]\!]} \max_{j \in [\![E_d]\!]} E_{q_i} \cdot E_{d_j}^T.
\end{equation}

For the retrieval task $\mathcal{R}$, the encoder returns the top-$k$ documents with the highest $Sim_{qd}$ scores as the most relevant to the query $q$. For computational efficiency, the search space of candidate documents can be proactively pruned by selecting the top-$K$ nearest clusters to the query, leveraging the memory.

\begin{algorithm}[t]
\LinesNumbered  
\caption{Update Memory Structure (Section \ref{sec:memory_update})}
\label{alg:memory_update}
\DontPrintSemicolon
\KwIn{Incoming stream $S_t$ at session $t$, Memory $M_t$ at session $t$, Assignment radius factor $\lambda$, Decaying radius factor $\gamma$}
\KwOut{Updated Memory $M_t$}
\tcc{\texttt{Cluster Initialization}}
\If{$t = 0$}{
    $C_0 \leftarrow ConstructInitialClusters(S_0)$\;
    \ForEach{$C \in C_0$}{
        $UpdateStatisticsAndPrototype(C)$\;
    }
}
\For{$t \in \{1, \cdots, T\}$}{
    \tcc{\texttt{Cluster Assignment}}
    \ForEach{$x \in S_t$}{
        $C \leftarrow FindNearestCluster(x, M_t)$\;
        $\mu_c, \sigma_c, p_c \leftarrow GetStatisticsAndPrototype(C)$\;
        \If{$SimDist(x, p_c) \leq \mu_c + \lambda \sigma_c$}{
            $Assign(x, C, M_t)$\;
        }
        \Else{
            $C \leftarrow AddNewCluster(x, M_t)$\;
        }
        $UpdateStatisticsAndPrototype(C)$\;
    }
    \tcc{\texttt{Cluster Maintenance}}
    \ForEach{$C \in M_t$}{
        $\mu_c, \sigma_c, p_c \leftarrow GetStatisticsAndPrototype(C)$\;
        \ForEach{$x \in C$}{
           \If{$SimDist(x, p_c) \geq \mu_c + \gamma \sigma_c$}{
              $M_t \leftarrow Remove(x, C, M_t)$\;
            }
        }
    }
}
\KwRet{$M_t$}
\end{algorithm}

\subsection{Memory Update}
\label{sec:memory_update}
\algname{} employs adaptive soft memory to implement the memory-based continual learning algorithm $\mathcal{A}$. To effectively represent the evolving topical distributions of documents and queries in a streaming setting, we adopt a streaming clustering for adaptive memory maintenance. This enables continual modeling of knowledge emergence and extinction over time. Specifically, \algname{} continuously assigns new queries and documents into topical clusters represented by the cluster prototypes whose sizes are regularized to preserve fine-grained semantics. Clusters are also managed sustainably with a decaying mechanism. Overall, Algorithm~\ref{alg:memory_update} outlines the main procedure for the memory update. First, initial clusters are constructed, and summary statistics and prototypes are computed for each cluster (Lines 1-4). Second, each new instance in the incoming stream is assigned to the nearest cluster or initiates a new cluster (Lines 6-13). Finally, instances beyond the threshold distance are removed, and statistics are updated (Lines 14–18).

\subsubsection{\textbf{\textit{Cluster Assignment with Regularized Prototype.}}} 
Typical clustering assigns a new data to the cluster with the nearest centroid. However, in the context of continual retrieval, the varying token lengths of queries and documents pose nontrivial challenges in representing cluster prototypes. While a simple mean pooling-based centroid represented in a single vector embedding is a straightforward solution, it compromises token-level semantics, which are essential in our fine-grained relevance estimation.

To fully exploit the token-level similarity introduced in Section \ref{sec:retrieval} while minimizing the high computational cost to address varying token lengths, we aggregate queries and documents in a cluster into a single prototype with a regularized, constant token length. Specifically, we adopt Random Projection Locality Sensitive Hashing (RP-LSH) to transform variable-length tokens into fixed-length vectors while maximally preserving the original semantic granularity. Unlike traditional LSH, which is typically tailored to Jaccard similarity for set-based data or Hamming distance for binary vectors, RP-LSH is well-suited for high-dimensional continuous embeddings and supports cosine similarity in the embedding space. Among alternatives such as Product Quantization, we choose RP-LSH for its balance between computational efficiency and representational fidelity. 

By projecting embeddings onto hash planes, RP-LSH maps semantically similar tokens to the same hash bucket, enabling efficient prototype construction with minimal loss of original semantic information. Let $d$ be the embedding dimension and $H$ the size of the RP-LSH hash space. The resulting prototype is an $H \times d$ matrix. The choice of $H$, controlled by the number of bits in the RP-LSH key, directly affects the trade-off between compression and expressiveness; i.e., smaller $H$ leads to information loss, while larger $H$ increases computational cost by approximating full token-wise comparisons.
For practical guidance on balancing the trade-off, we provide a theoretical analysis on the choice of LSH bit size:

\begin{theorem}
\vspace{-0.1cm}
\label{theorem:sufficiency} 
{\sc (Sufficient LSH bitsize)} When generating prototypes from \( M \) token embeddings, the sufficient number of LSH bits is determined as $ \log_2 \left( \frac{8 \ln M}{\varepsilon^2} \right)$ at the optimal distortion rate $\varepsilon = \frac{1}{3\sqrt{e}}$.
\end{theorem}
\vspace{-0.3cm}
\begin{proof}
A benefit function is defined to trade off the gain in accuracy against the computational cost. The Johnson-Lindenstrauss (JL) lemma~\cite{JL_lemma} provides a cost model in terms of distortion $\varepsilon$, while the approximation quality imposes a lower bound on $\varepsilon$. Maximizing the benefit within this feasible range yields the optimal distortion rate $ \varepsilon^* = \frac{1}{3\sqrt{e}}$. Using the JL lemma, we derive the minimal bit size for LSH at $\varepsilon^*$. See Appendix \ref{apx:full_proof} for the full proof.
\end{proof}

For example, in our evaluation setting with the LOTTE dataset~\cite{colbertv2}, each session includes approximately 2,430 queries (with an average of 9 tokens) and 500,000 documents (with an average of 159 tokens), resulting in up to 80 million token embeddings per session, each with 768 dimensions by BERT~\cite{bert}. Then, according to Theorem~\ref{theorem:sufficiency}, to maintain an acceptable distortion rate of $\varepsilon = \frac{1}{3\sqrt{e}} \approx 0.2$, the sufficient number of RP-LSH bits is given by:
\begin{equation}
\label{eq:k_bit}
h \geq \lceil \log_2\left(\frac{8 \ln(8\times10^7)}{(0.2)^2}\right) \rceil \approx 12.
\end{equation}
Thus, we employ a 12-bit RP-LSH, resulting in $H = 2^{12} = 4{,}096$ hash buckets. Each bucket is initialized with a zero vector, and the aggregated embeddings within it are normalized, collectively serving as the cluster prototype in a vector of size (4,096, 768). Assuming that the approximate number of clusters in LoTTE remains around 12, \algname{} computes similarities on 4,096 fixed-length arrays derived from the LSH prototypes instead of full token embeddings. This dramatically reduces the number of token-pairwise operations per session from $1.7 \times 10^{12}$ (i.e., $(2,430 \times 9) \times (500,000 \times 159)$ token pairs) to $1.1 \times 10^{9}$ (i.e., $(2,430 \times 9) \times (4,096 \times 12)$ token–prototype pairs), yielding roughly a $1.6 \times 10^{3}$times reduction in token-level computations.

Each new query $q$ or document $d$, it is assigned to the nearest cluster if its distance to the prototype is within $\mu + \lambda \sigma$, where $\lambda$ is a tunable assignment factor; otherwise, a new cluster is initialized.

\subsubsection{\textbf{\textit{Lightweight Cluster Maintenance.}}} 
For a query $q$ or a document $d$, we derive its distance $SimDist$ to a cluster prototype $p$ from the token-level similarity (e.g., $SimDist_{pd} = L - Sim_{pd}$ where $L$ is the maximum token length of the encoder). These distances serve as the primary metric for  cluster maintenance. 

\algname{} summarizes each cluster by the compact triplet $\langle N, LS, SS \rangle$, similar to the cluster feature vector in BIRCH~\cite{birch_cfv}. In \algname{}, however, we track only the distance summaries, which are even more compact than the original embedding summaries in BIRCH. Specifically, $N$ denotes the number of instances in the cluster, $LS$ is the linear sum of distances to the prototype, and $SS$ is the sum of squared distances to the prototype. This cluster summary is sufficient to compute key cluster statistics, such as the mean and standard deviation of distances to the prototype, and also allows efficient incremental updates in an additive manner.

Retaining all past queries and documents from unbounded streaming corpora is not feasible in practical scenarios and can degrade continual learning performance due to the accumulation of outdated knowledge. To address this, we adopt a radius-decaying cluster maintenance policy that selectively preserves representative documents in clusters. At the end of each session, we retain only the documents whose distance to their cluster prototype falls below $\mu + \gamma \sigma$, where $\gamma$ is a tunable decaying factor. For queries, we perform random sampling proportional to the number of retained documents in each cluster. The assignment factor $\lambda$ determines whether new samples are assigned to existing clusters, while the decaying factor $\gamma$ prunes semantically less important samples in the existing clusters at the end of each session. This regulates a forgetting mechanism to explicitly control document accumulation, improving scalability and reducing computational overhead without sacrificing performance. In addition, \algname{} can adopt a multi-stage retrieval-and-sampling pipeline; e.g., BM25-based pre-filtering followed by candidate subsampling, enabling an even more lightweight memory construction.




\subsection{Encoder Update}
\label{sec:encoder_update}
To train the encoder to learn an effective embedding space for matching relevant queries and documents, we employ self-supervised contrastive learning with the aid of the soft memory. \algname{} samples a diverse set of queries that represent each cluster, and leverages inter-topic semantic relationships to select both positive and negative documents for each query. The pseudo-codes are provided in Appendix~\ref{apx:pseudo_code}. The structural properties of the soft memory help reduce the search space, enabling efficient training without exhaustive processing of all queries and documents. 

\subsubsection{\textbf{\textit{Query Selection.}}}
To construct a training sample that effectively reflects the entire knowledge space in the memory, we propose a query sampling strategy by stratified sampling and coreset selection~\cite{coreset}. A relevant approach, topic-aware sampling~\cite{topic_aware_sampling}, has shown effectiveness in a static IR setting. While it randomly samples queries from a limited subset of clusters, \algname{} considers the full cluster distribution and explicitly selects an optimal query set by maximizing coverage over the entire cluster space. 

Specifically, we sample queries from each cluster $C_i$ in proportion to its size, to mitigate bias toward large clusters. The number $N_i$ of queries selected from cluster $C_i$ is defined as $N_i = N \cdot \frac{|D_i|}{|D|}$, where $N$ is the total number of queries to sample and $D$ is the entire document set in the memory. Let $Q_i$ and $D_i$ denote the sets of queries and documents, respectively, in cluster $C_i$. Each query $q \in Q_i$ is associated with a set $D_{q} \subseteq D_i$ of top-$m$ closed documents, where $m = \frac{|D_i|}{|Q_i|}$. \algname{} aims to find the optimal subset of queries of which document coverages are minimum-overlapping; i.e., the union of all $D_{q}$ maximally covers $D_i$. Given the candidate query-documents pairs $U = \{ (q, D_q) \mid q \in Q_i \}$, a seed pair $u$ is randomly chosen. Then, \algname{} iteratively selects the next query $q^*$ in $u^*$ that maximizes document coverage and minimizes redundancy:
\begin{equation}
\label{eq:max_coverage_simplified}
\begin{aligned}
q^* &= \arg\min_{q \in u^*} 
\left| D_q \cap \bigcup_{q' \in u} D_{q'} \right|, \quad \text{where} \\
u^* &= \left\{ q \in U \setminus u \;\middle|\; \left| \bigcup_{q' \in u \cup \{q\}} D_{q'} \right| = \max_{q'' \in U \setminus u} \left| \bigcup_{q' \in u \cup \{q''\}} D_{q'} \right| \right\}.
\end{aligned}
\end{equation}


\noindent This process is repeated until $N_i$ queries are selected for each cluster, to construct the final training set $U = \bigcup u$.

\subsubsection{\textbf{\textit{Document Selection.}}}
When searching for positive and negative documents for each query, exhaustively scanning the entire document collection is inefficient. Therefore, we restrict the search space to the top-$K$ nearest clusters retrieved for each query. Considering false positives resulting from the approximate nature of RP-LSH, we select the most similar document as the positive sample and the least similar documents as negative samples, ensuring that they are clearly distinguishable. As a result, the representation becomes increasingly aligned with documents from the same topic as the query—typically found in the top-1 cluster—while diverging from semantically similar documents belonging to different topics, which are located in the top-2 to top-$(K{-}1)$ clusters. Let $C_K(q)$ be the set of documents in the top-$K$ clusters retrieved for query $q$. For each query, we construct a training document set $T_q = \{ d^+, d_1^-, \dots, d_{k-1}^- \}$, where $d^+$ is the most similar document and $\{ d_j^- \}_{j=1}^{k-1}$ are the least similar documents in $C_K(q)$ based on token-level similarity $Sim_{qd}$:
\begin{equation}
\label{eq:doc_sampling}
\begin{aligned}
T_q &= \{ d^+ \} \cup \{ d_j^- \}_{j=1}^{k-1}, \\
\text{where} \quad
d^+ &= \arg\max_{d \in C_K(q)} Sim_{qd}, \\
\{ d_j^- \}_{j=1}^{k-1} &= \arg\min\nolimits^{k-1}_{d \in C_K(q) \setminus \{ d^+ \}} Sim_{qd}.
\end{aligned}
\end{equation}

\subsubsection{\textbf{\textit{Training Objective.}}} 
For each sampled query $q \in U$ and the associated documents $T_q$, \algname{} treats the corresponding positive and negative documents as pseudo-labels and trains the encoder to assign higher similarity to the positive pairs $(q, d^+)$ than to negative pairs $(q, d^-)$ through a contrastive objective:

\begin{equation}
\label{eq:contrastive_loss}
\mathcal{L} = -\log \frac{\exp\left(  sim(q, d^+) / \tau \right)}{\sum\limits_{d \in T_q} \exp\left( sim(q, d) / \tau \right)}, \quad \forall q \in U.
\end{equation}


%% file: tex/05-Experiments.tex
\begin{table*}[!ht]
\centering
\setlength{\tabcolsep}{3pt}
\renewcommand{\arraystretch}{0.95} 
\caption{Overall performance on LoTTE (bold: best in unsupervised; underlines: best in all settings including supervised$^*$.) }
\vspace{-0.30cm}
\begin{adjustbox}{max width=\textwidth}
\label{tb:overall_lotte}
\begin{tabular}{l|cc|cc|cc|cc|cc|cc|cc|cc|cc|cc|cc}
\toprule
Session & \multicolumn{2}{c|}{0} & \multicolumn{2}{c|}{1} & \multicolumn{2}{c|}{2} & \multicolumn{2}{c|}{3} & \multicolumn{2}{c|}{4} & \multicolumn{2}{c|}{5} & \multicolumn{2}{c|}{6} & \multicolumn{2}{c|}{7} & \multicolumn{2}{c|}{8} & \multicolumn{2}{c|}{9} & \multicolumn{2}{c}{Avg} \\
 & S@5 & R@10 & S@5 & R@10 & S@5 & R@10 & S@5 & R@10 & S@5 & R@10 & S@5 & R@10 & S@5 & R@10 & S@5 & R@10 & S@5 & R@10 & S@5 & R@10 & S@5 & R@10 \\
\hline
\textbf{ColBERT\textsuperscript{+}*} & 1.7 & 0.3 & 8.5 & 4.6 & 15.2 & 5.2 & 16.7 & 7.1 & 18.9& 8.8 & 20.7 & 10.4 & 25.2 & 10.5 & 33.2 & 14.3 & 39.4 & 20.9 & 32.8 & 17.9& 21.23 &10.00\\
\textbf{ER*} & 36.1 & 17.6 & 45.9 & 25.4 & 38.
5& 19.8 & 47.4 & 27.2 & 53.3 & 28.4 & 41.1 & 21.4 & 45.9 & 23.8& 40.0 & 21.6 & 58.9 & 35.7 & \underline{52.2} & \underline{29.7} & 45.93 & 25.06 \\
\textbf{MIR*} & 36.7 & \underline{18.0} & 46.3 & \underline{25.7} & 36.3 & 15.7 & 45.9 & 26.7 & 51.5 & 28.3 & 42.6 & 21.4 & 44.8 & 23.6 & 42.6 & 21.8 & 61.1 & 40.1 & 48.3 & 27.4 & 45.61 & 24.87 \\
\textbf{GSS*} & 36.1 & 15.7 & 42.2 & 23.6 & 39.3 & 18.6 & 48.1 & 27.6& 49.3 & 26.1 & \underline{44.8} & \underline{23.1} & 45.9 & 23.9 & 37.9 & 20.9 & 61.1 & 37.5 & 43.3 & 24.4 & 44.80 & 24.14 \\
\textbf{OCS*} & 36.1 & 16.7 & 42.6 & 22.3 & 37.8 & 18.6 & 46.7 & 25.8 & 47.8 & 24.3 & 41.1 & 21.8 & 44.8 & 24.1 & 40.0 & 21.6 & 59.4 & 36.1 & 44.4 & 26.3 & 44.07 & 23.76\\
\textbf{L2R*} & 17.8 & 7.2 & 34.4 & 16.7 & 27.8 & 12.1 & 38.9 & 21.2 &38.1 & 17.9&  28.1 & 13.9 & 33.0 & 16.9 & 27.2 & 12.2 & 48.3 & 26.2 & 34.4 & 18.6 & 32.80 & 16.29 \\
\midrule
\rowcolor{gray!5}
\textbf{BM25} & 26.1
& 12.4 & 41.1 & 17.7 & 45.6 & 20.4 & 42.2 & 22.4 & 44.8 & 23.3 &34.8 &17.1 & 34.4 & 15.1 & \underline{\textbf{48.1}} & \underline{\textbf{22.4}} & 50.0 & 27.0 & 40.6 & 21.0 & 40.77 & 19.88 \\
\rowcolor{gray!5}
\textbf{ColBERT\textsuperscript{+}} & 0.0 & 0.1 & 1.1 & 0.2 & 0.7 & 0.2 & 1.9 & 0.5 & 0.7 & 0.1 & 0.4 & 0.2 & 4.4 & 2.1 & 3.4 &1.9 & 11.1 & 4.6 & 13.3 & 10.5 & 3.70 & 2.04 \\
\rowcolor{gray!5}
\textbf{ER} & 15.0 & 5.9 & 25.9 & 11.0 & 16.3 & 7.1 & 25.2 & 13.8 & 23.7 & 10.6 & 17.4 & 7.9 & 20.7 & 11.1 & 21.3 & 9.6 & 36.7 & 19.6 & 23.9 & 11.0 &22.61 &10.76\\
\rowcolor{gray!5}
\textbf{MIR}& 14.4 & 5.6 & 21.9 & 9.0 & 19.3 & 6.9 & 27.4 & 14.9 & 21.5 & 9.5 & 12.2 & 6.8 &  19.3 & 9.4 & 12.8 & 5.8 & 32.8 & 14.7 & 13.9 &7.1 & 19.55 &8.97 \\
\rowcolor{gray!5}
\textbf{GSS} & 13.9 & 5.8 & 25.9 & 11.3 & 22.2 & 9.6 & 28.5 & 16.0 & 23.7 & 10.1 & 16.3 & 6.7 & 20.7 & 9.2 & 15.7 & 6.1 & 30.6 & 15.8 & 13.3 & 8.0 & 21.08 & 9.86 \\
\rowcolor{gray!5}
\textbf{OCS}& 14.4 & 5.7 & 22.2 & 10.4 & 15.2 & 7.3 & 27.0 & 14.6 & 25.2 & 11.1 & 14.1 & 6.4 & 20.7 & 8.8 & 14.5& 7.1 & 33.9 & 16.0 & 16.1 & 9.0 & 20.33 & 9.64 \\
\rowcolor{gray!5}
\textbf{L2R}& 15.0 & 5.9 & 23.0 & 10.1 & 14.1 & 6.2 & 26.3 & 15.4 & 26.3 & 11.7 & 16.7 & 7.2 & 20.7 &11.1 & 17.9 & 7.1 & 35.6 & 19.6 & 14.4 & 7.0 &  22.61 & 9.92 \\
\hdashline
\rowcolor{gray!15}
\textbf{\algname} & \underline{\textbf{37.2}} & \textbf{16.7} & \underline{\textbf{47.4}} & \textbf{24.8} & \underline{\textbf{47.8}} & \underline{\textbf{23.6}} & \underline{\textbf{57.8}} & \underline{\textbf{32.8}} & \underline{\textbf{58.5}} & \underline{\textbf{33.2}} & \textbf{43.7} & \textbf{22.5} & \textbf{35.0} & \underline{\textbf{24.5}} & 45.2 & 21.5 & \underline{\textbf{66.7}} & \underline{\textbf{45.4}} & \textbf{46.7} & \textbf{26.1} & \underline{\textbf{48.60}} & \underline{\textbf{27.11}} \\
\bottomrule
\end{tabular}
\end{adjustbox}
\end{table*}

\begin{table*}[t]
\centering
\setlength{\tabcolsep}{3pt}
\renewcommand{\arraystretch}{0.95} 
\caption{Overall performance on MSMARCO (bold: best in unsupervised; underlines: best in all settings including supervised$^*$.) }
\vspace{-0.30cm}
\begin{adjustbox}{max width=\textwidth}
\label{tb:overall_msmarco}
\begin{tabular}{l|cc|cc|cc|cc|cc|cc|cc|cc|cc|cc|cc}
\toprule
Session & \multicolumn{2}{c|}{0} & \multicolumn{2}{c|}{1} & \multicolumn{2}{c|}{2} & \multicolumn{2}{c|}{3} & \multicolumn{2}{c|}{4} & \multicolumn{2}{c|}{5} & \multicolumn{2}{c|}{6} & \multicolumn{2}{c|}{7} & \multicolumn{2}{c|}{8} & \multicolumn{2}{c|}{9} & \multicolumn{2}{c}{Avg} \\
 & S@5 & R@10 & S@5 & R@10 & S@5 & R@10 & S@5 & R@10 & S@5 & R@10 & S@5 & R@10 & S@5 & R@10 & S@5 & R@10 & S@5 & R@10 & S@5 & R@10 & S@5 & R@10\\
\hline
\textbf{ColBERT\textsuperscript{+}*} & 2.8 & 6.1 & 9.6 & 13.2 & 19.3 & 23.8 & 20.4 & 23.5 & 21.9 & 26.9 & 68.1 & 75.6 & 87.4 & 89.3 & 71.1 & 77.7 & 84.8 & 89.3 & 92.2 & 93.2 & 47.76 & 51.86\\
\textbf{ER*}  & 56.1 & 65.0 & 64.8 & 71.4 & 66.7 & 75.2 & 65.6 & 70.6 & 61.5 & 66.8 & 89.3 & \underline{91.5} & \underline{89.3} & 79.4 & 90.4 & 92.8 & 90.4 & 93.1 & \underline{88.9} & \underline{91.4} & 76.30 & 80.85 \\
\textbf{MIR*} & \underline{60.6} & \underline{70.8} & \underline{69.3} & \underline{74.7} & 63.7 & 73.0 & 63.0 & 66.4 & 61.9  & 67.5 & 88.1 & 90.9 & \underline{89.3} & \underline{91.5} & \underline{91.9} & \underline{93.1} & 85.9 & 90.2 & 90.0 & 92.7 & \underline{76.37} & \underline{81.08} \\
\textbf{GSS*}  & 58.3 & 68.6 & 64.4 & 70.2 & 61.1 & 69.8 & 66.7 & 72.0 & \underline{65.9} & \underline{71.9} & \underline{90.0} & 91.4 & 88.1 & 90.4 & 88.9 & 91.5 & 86.7 & 90.6 & 88.1 & 89.4 & 75.82 &80.63 \\
\textbf{OCS*}  & 55.6 & 61.7 & 67.0 & 71.1 & 64.1 & 72.2 & 57.8 & 64.9 & 57.0 & 65.1 & 83.3 & 87.7 & 84.4 & 88 & 86.7 & 90.3 & 86.3 &87.1 & 88.5 & 90.5 & 73.07 & 77.86\\
\textbf{L2R*}  & 35.0 & 40.0 & 44.4 & 51.9 & 47.4 & 58.1 & 46.7 & 55.7 & 48.9 & 55.1 & 78.1 & 81.7 & 81.9 & 85.7 & 82.6 & 85.7 & 81.1 & 86.5 & 83.7 & 85.4 & 62.98 & 68.58 \\
\midrule
\rowcolor{gray!5}
\textbf{BM25} & 25.0 & 26.9 & 35.6 & 40.6 & 41.5 & 47.5 & 49.3 & 56.5 & 48.1 & 52.4 & 51.5 & 55.0 & 52.6 & 57.2 & 60.4 & 64.1 & 74.1 & 76.5 & 68.1 & 68.7 & 50.62 & 54.54 \\
\rowcolor{gray!5}
\textbf{ColBERT\textsuperscript{+}} & 0.0 & 0.0 & 1.9 & 2.6 & 5.9 & 8.6 & 3.7 & 4.8 & 0.7 & 1.6 & 18.1 & 24.0 & 22.6 & 29.3 & 32.2 & 38.5 & 30.4 & 35.6 & 23.7 & 29.3 &13.92 &17.43 \\
\rowcolor{gray!5}
\textbf{ER} & 17.2 & 22.5 & 23.3 & 27.8 & 34.4 & 41.5 & 27.0 & 33.8 & 17.4 & 19.4 & 50.0 & 57.3 & 56.7 & 63.0 & 67.8 & 72.3 & 66.3 & 71.1 & 55.9 & 61.9 &41.60 &47.06 \\
\rowcolor{gray!5}
\textbf{MIR}  & 18.9 & 23.3 & 25.6 & 30.0 & 34.8 & 41.3 & 30.4 & 35.9 & 18.9 & 21.2 & 56.7 & 64.3 & 60.0 & 66.5 & 70.0 & 73.7 & 65.2 & 73.7 & 60.7 & 67.4 &44.12 &49.73 \\
\rowcolor{gray!5}
\textbf{GSS}  & 17.2 & 21.9 & 20.0 & 26.0 & 29.6 & 35.6 & 20.4 & 25.3 & 12.2 & 19.6 & 47.0 & 51.4 & 46.3 & 54.1 & 60.7 & 65.9 & 63.7 & 69.6 & 58.9 & 62.2 & 37.60 & 43.16 \\
\rowcolor{gray!5}
\textbf{OCS}  & 16.7 & 24.2 & 24.8 & 29.1 & 33.3 & 41.3 & 25.6 & 34.3 & 21.9 & 25.6 & 57.0 & 64.2 & 58.5 & 64.4 & 67.0 & 70.2 & 62.6 & 70.2 & 58.9 & 63.3 & 42.63 & 48.68 \\
\rowcolor{gray!5}
\textbf{L2R}  & 20.0 & 24.4 & 21.5 &25.8 & 27.8 & 34.4 & 19.6 & 24.6 & 15.6 & 20.2 & 45.9 & 52.2 & 48.1 & 56.1 & 61.5 & 66.0 & 56.3 & 62.8 & 51.9 & 59.5 & 36.82 & 42.60 \\
\hdashline
\rowcolor{gray!15}
\textbf{\algname{}}& \textbf{57.2} & \textbf{65.0} & \textbf{57.4} & \textbf{65.3} & \underline{\textbf{65.9}} & \underline{\textbf{75.1}} & \underline{\textbf{68.9}} & \underline{\textbf{76.7}} & \textbf{63.3} & \textbf{69.3} & \textbf{78.9} & \textbf{81.0} & \textbf{78.1} & \textbf{73.5} & \textbf{90.4} & \textbf{92.4} & \underline{\textbf{92.6}} & \underline{\textbf{94.4}} & \textbf{84.1} & \textbf{86.2} & \textbf{73.68} & \textbf{77.89} \\
\bottomrule
\end{tabular}
\end{adjustbox}
\end{table*}

\section{Experiments}
We evaluate the efficacy and efficiency of \algname{} to answer the following questions.
\begin{itemize}[leftmargin=9pt, noitemsep]
    \item How does the proposed framework perform in general relative to the baselines in real-world datasets with the two evaluation protocols $\mathcal{R_{shared}^*}$ and $\mathcal{R_{disjoint}^*}$? (Section \ref{sec:overall_performance})
    \item Are the main techniques employed in our framework effective? (Section \ref{sec:ablation_study}).
    \item How robust is the performance of the proposed method to changes in its key parameters? (Section \ref{sec:param_sensitivity})
\end{itemize}

\subsection{Experiment Setup}
\noindent\textbf{\uline{\textit{Datasets.}}}
To effectively model the dynamics of evolving data distributions, we conduct experiments on two real-world benchmark datasets widely used for continual retrieval~\cite{l2r, continual_msmarco}: LoTTE~\cite{colbertv2} and MSMARCO~\cite{modified_msmarco} with the queries clustered by topic from the original dataset~\cite{msmarco}. LoTTE includes five domains (writing, recreation, science, technology, and lifestyle) from StackExchange questions and answers. MSMARCO includes five domains (Names and Public Figures, Dated Events, Pricing/Units, Medical Treatments and Biology/Physics.) from Bing questions and answers. Each dataset is simulated in 10 streaming sessions with partially overlapping topics following the convention in continual learning~\cite{clsurvey, dark_bbcl}. Based on the two evaluation protocols formalized in Section \ref{sec:problem_setting}, the training and evaluation query sets are disjoint, and the training and evaluation document sets are either separated or shared. Details of datasets and sessions are given in Appendix \ref{apx:dataset}. 

\noindent\textbf{\uline{\textit{Baselines and Implementation.}}}
For Sparse Retrieval, we use BM25, and for Dense Retrieval, we use ColBERT\textsuperscript{+}, a continual learning variant of ColBERT. For Memory-based Continual Learning (MCL), we evaluate five methods: Experience Replay (ER), Maximally Interfered Retrieval (MIR), Gradient-based Sample Selection (GSS), Online Coreset Selection (OCS), and L2R. We evaluate these baselines and \algname{} in an unsupervised setting, with their supervised variants denoted by an asterisk(*). Implementation details for all baselines and \algname{} are provided in Appendix \ref{apx:implementation}.

\noindent\textbf{\uline{\textit{Evaluation Metrics.}}}
Two standard metrics, Success@5 (S@5) and Recall@10 (R@10), are used. Success@$k$ indicates whether at least one ground-truth positive document is retrieved within the top-$k$ results for a given query. Recall@$k$ measures the proportion of all relevant documents that appear within the top-$k$ retrieved results.

\subsection{\textbf{Overall 
Performance}}
\label{sec:overall_performance}
\noindent\textbf{\uline{\textit{Comparison with Baselines.}}}
As shown in Tables~\ref{tb:overall_lotte} and \ref{tb:overall_msmarco}, under $\mathcal{R_{shared}^*}$, \algname{} consistently outperforms all baselines in the unsupervised setting across all sessions on both datasets. On average, it surpasses BM25—the strongest baseline—by 19.21\%/36.37\% (Success@5/Recall@10) on LoTTE and 46.19\%/42.81\% on MSMARCO. In the supervised setting, \algname{} achieves the highest average performance on LoTTE, outperforming even supervised MCL methods and ColBERT\textsuperscript{+}. Compared to ER*, the best supervised baseline, it improves Success@5 and Recall@10 by 5.81\% and 8.18\%, respectively. On MSMARCO, it also exceeds the average performance of supervised OCS, L2R, and ColBERT\textsuperscript{+}. These results highlight \algname{} achieves performance on par with state-of-the-art supervised methods, despite using no supervision. As shown in Table \ref{tb:separated_evaluation}, the disjoint evaluation $\mathcal{R_{disjoint}^*}$ showed similar results.

\noindent\textbf{\uline{\textit{Sparse Retrieval vs. Dense Retrieval.}}}
In the label-free setting, BM25 outperformed MCL baselines and ColBERT\textsuperscript{+}, except for our method, highlighting the robustness of sparse retrieval under domain shifts without learning. Notably, ColBERT\textsuperscript{+} showed only marginal gains in early evaluations, due to several architectural and training constraints. First, it introduces special tokens, increasing the number of token types to be learned. Although it shares the same backbone as the baselines, additional linear layers increase the number of trainable parameters and overall learning complexity. Furthermore, ColBERT\textsuperscript{+} employs a late interaction mechanism over compressed low-dimensional representations, which typically require extensive training. Also, the streaming simulation setting with only one training epoch causes the model to underfit. Its performance improves significantly in later sessions, but poor early-stage performance lowers the overall average.

\noindent\textbf{\uline{\textit{Analysis of Continual Learning Methods.}}}
\algname{} achieved the best performance among dense retrievers on LoTTE in both supervised and unsupervised settings, as well as on MSMARCO in the unsupervised setting, whereas MIR achieved the highest performance on MSMARCO in the supervised setting. Although supervised ER and MIR are relatively simple methods, they demonstrated strong performance, likely due to their ability to sample diverse training instances across distributions via random sampling. In contrast, methods like OCS, GSS, and L2R tend to select samples similar to existing positives, which helps capture intra-distribution relationships but limits diversity across domains.

\noindent\textbf{\uline{\textit{Analysis between Datasets.}}}
We observed that \algname{} achieved a larger performance gain over baselines on LoTTE compared to MSMARCO. This is likely due to the use of ground-truth domain labels in LoTTE, whereas MSMARCO relies on pseudo-domain labels generated via clustering, which may introduce noise.

\begin{table}[t]
\centering
\small
\caption{Overall performance with disjoint evaluation protocol $\mathcal{R_{disjoint}^*}$ (bold: best in unsupervised; underlines: best in all settings including supervised$^*$.) }
\vspace{-0.3cm}
\label{tb:separated_evaluation}
\begin{tabular}{l|cc|cc}
\toprule
 & \multicolumn{2}{c}{\textbf{LoTTE}} & \multicolumn{2}{c}{\textbf{MSMARCO}}\\
 & S@5 & R@10 & S@5 & R@10 \\
\hline
\textbf{ColBERT\textsuperscript{+}*}   & 42.13 & 23.30 & 78.29 & 82.95\\
\textbf{ER*}    & 67.59 & 43.24 & \underline{96.32} & \underline{97.69}\\
\textbf{MIR*}   & 67.33 & 42.57& 95.66 & 96.90\\
\textbf{GSS*}   & 68.15 & 43.57 & 95.77 & 94.49\\
\textbf{OCS*}   & 66.77 & 42.75 & 95.98 & 97.14\\
\textbf{L2R*}   & 55.06 & 31.67 & 91.00 & 94.04\\
\midrule
\rowcolor{gray!5}
\textbf{BM25}  & 59.89 & 33.20 & 72.79 & 75.49 \\
\rowcolor{gray!5}
\textbf{ColBERT\textsuperscript{+}}  & 4.11 & 1.43 & 41.19 & 51.50\\
\rowcolor{gray!5}
\textbf{ER}   & 40.51 & 25.73 & 75.92 & 81.86\\
\rowcolor{gray!5}
\textbf{MIR}   & 31.83 & 16.69 & 76.44 & 82.32\\
\rowcolor{gray!5}
\textbf{GSS}   & 37.25 & 20.53 & 75.33 & 82.29\\
\rowcolor{gray!5}
\textbf{OCS}   & 39.43 & 22.24 & 73.52 & 79.89\\
\rowcolor{gray!5}
\textbf{L2R}   & 41.57 & 23.60 & 75.05 & 81.44\\
\hdashline
\rowcolor{gray!15}
\textbf{\algname{}}  & \underline{\textbf{68.20}} & \underline{\textbf{43.57}} & \textbf{93.15} & \textbf{95.15} \\
\bottomrule
\end{tabular}
\end{table}

\subsection{\textbf{Ablation Study}}
\label{sec:ablation_study}
We evaluate the efficacy of the three main components of \algname{}:
\begin{itemize}[leftmargin=9pt, noitemsep]
    \item \textbf{w/o fine-grained similarity} does not consider token-level similarity and regularized prototype. All queries and documents are represented as mean-pooled vectors, cluster prototypes are defined as mean-pooled centroids, and cosine similarity is used.
    \item \textbf{w/o update encoder} does not consider training encoder. Evaluation is performed based on token-level similarity.
    \item \textbf{w/o soft memory} does not consider soft memory and performs naive incremental learning without clustering. For each query, the document with the highest cosine similarity across the entire corpus is selected as the positive, while the least similar documents are chosen as negatives.
\end{itemize}

As shown in Table~\ref{tb:ablation_study}, on both datasets, the full method with all components consistently achieved the best performance across most sessions. This clearly demonstrates that all components contribute jointly to the overall performance. The largest performance drop was observed when removing fine-grained similarity and the regularized prototype (an average drop of 44.93\% in Success@5 and 42.72\% in Recall@10), indicating that effectively leveraging fine-grained semantics plays a crucial role in performance under unsupervised settings. This was followed by the contributions of removing soft memory (which resulted in an average drop of 18.08\% in Success@5 and 20.57\% in Recall@10) and removing the update encoder (which resulted in an average drop of 8.24\% in Success@5 and 9.83\% in Recall@10), in that order. Notably, performing incremental learning without soft memory resulted in greater performance degradation compared to not training at all. This suggests the necessity of both high-quality sampling through soft memory and continual learning. The impact of the update encoder is relatively smaller compared to other components, yet its removal still causes a noticeable performance drop, indicating that encoder updates are necessary for adapting to new data distributions.

\begin{table}[!t]
\centering
\small
\caption{Ablation study results.}
\vspace{-0.3cm}
\label{tb:ablation_study}
\begin{tabular}{l|cc|cc}
\toprule
 & \multicolumn{2}{c|}{\textbf{LoTTE}} & \multicolumn{2}{c}{\textbf{MSMARCO}}\\
 & S@5 & R@10 & S@5 & R@10 \\
\midrule
\textbf{\algname{}} & \textbf{48.60} & \textbf{27.11} & \textbf{73.68} & \textbf{77.89} \\
\textbf{- w/o fine-grained similarity} & 27.23 & 13.33 & 44.30 & 50.94 \\
\textbf{- w/o update encoder} & 46.07 & 24.26 & 65.38 & 70.77 \\
\textbf{- w/o soft memory} & 38.08 & 19.45 & 62.77 & 67.86 \\
\bottomrule
\end{tabular}
\vspace{-0.45cm}
\end{table}

\begin{table}[!t]
\small
\centering
\caption{Performance with varying LSH bit sizes and number of initial clusters on LoTTE and MSMARCO data sets.}
\label{tb:parameter_sensitivity}
\vspace{-0.3cm}
\begin{tabular}{c|c|cc|cc}
\toprule
\multirow{2}{*}{\textbf{Parameter}}
 & \multirow{2}{*}{\textbf{Value}} & \multicolumn{2}{c|}{\textbf{LoTTE}} & \multicolumn{2}{c}{\textbf{MSMARCO}} \\
 & & S@5 & R@10 & S@5 & R@10 \\
\midrule
\multirow{3}{*}{LSH bit size}
 & 0   & 45.08 & 23.44 & 65.04 & 70.31 \\
 & 6   & 48.32 & 26.07 & 70.19 & 75.42 \\
 & 12  & 48.60 & 27.11 & 73.68 & 77.89 \\
\midrule
\multirow{3}{*}{Initial clusters}
 & 3   & 32.60 & 16.32 & 75.11 & 79.91 \\
 & 12  & 48.60 & 27.11 & 73.68 & 77.89 \\
 & 48  & 48.87 & 25.59 & 69.77 & 74.99 \\
\bottomrule
\end{tabular}
\vspace{-0.45cm}
\end{table}

\subsection{Hyperparameter Sensitivity Analysis}
\label{sec:param_sensitivity}
We evaluate the performance of \algname{} under variations of its two key hyperparameters: LSH bit size (0, 6, 12) and the number of initial clusters (3, 12, 48). The further analysis of the assignment factor $\lambda$ and the decaying factor $\gamma$ is provided in Appendix \ref{apx:factor_analysis}.


As shown in Table~\ref{tb:parameter_sensitivity}, increasing the LSH bit size leads to a proportional improvement in retrieval performance. Utilizing 4,096 embeddings as prototypes yields an average gain of 10.54\% in Success@5 and 13.22\% in Recall@10, compared to using a single embedding as a prototype. This suggests that finer prototype granularity enables clusters to capture semantic distinctions more effectively. Regarding the number of clusters, the optimal configuration differs across datasets: LoTTE achieves the best performance with 12 clusters, whereas MSMARCO performs best with 3. Although both datasets span five domains, LoTTE includes 12 explicitly defined subtopics, while MSMARCO lacks a clear subtopic structure. This indicates that clustering functions not merely as a partitioning mechanism, but rather as a topic-aware abstraction of the data. The observed discrepancy in optimal cluster sizes can likely be attributed to differences in the underlying topic hierarchies. 

Notably, \algname{} consistently outperforms unsupervised baselines across varying parameter configurations on both datasets. This observation suggests that \algname{} is suitable for practical deployment, as it does not rely on extensive fine-tuning of its main hyperparameters. Moreover, the robustness of performance across different LSH bit sizes indicates that the proposed prototype regularization can prioritize resource efficiency through higher compression without degrading performance.

%% file: tex/06-Conclusion.tex
\section{Conclusion and Future Work}
In this work, we propose \algname{}, an unsupervised continual learning framework for dynamic information retrieval in which query and document distributions evolve over time. \algname{} integrates fine-grained token-level similarity with a clustering-based soft memory, enabling efficient encoder updates through selective query–document sampling from the memory. Experimental results on two benchmark datasets demonstrate substantial improvements in retrieval accuracy over existing baselines.

Toward more practical applicability, the evaluation can be extended along three axes: (i) broader task coverage beyond question answering (e.g., summarization), (ii) encompassing both recurring and non-recurring domain dynamics and leveraging corpora with explicit temporal metadata (e.g., timestamps) of multifaceted distribution drift, and (iii) more comprehensive evaluation metrics that jointly capture retrieval quality (e.g., ranking), retention of previously acquired knowledge, and acquisition of new information.

Furthermore, this work opens promising directions for agent-based AI systems: (i) extending the soft memory into a hierarchical representation could support multi-level sampling, thereby improving robustness to complex non-stationary shifts. (ii) the soft memory could evolve into an expandable knowledge base for agentic retrieval systems, enabling the ingestion of new documents, verification of evidence with temporal provenance, and prioritization of high-utility information under constrained context budgets.

%% file: tex/07-Appendix.tex
\clearpage
\appendix

\section{Appendix}

\subsection{Full Proof of Theorem \ref{theorem:sufficiency}}
\label{apx:full_proof}
\begin{proof}
We aim to find the optimal distortion rate \( \varepsilon \) by maximizing the benefit function \( B(\varepsilon) = \frac{P(\varepsilon)}{K(\varepsilon)} \), where \( P(\varepsilon) \) represents the accuracy gain and \( K(\varepsilon) \) represents the computational cost. The gain function \( P(\varepsilon) \) is modeled with two assumptions: (i) as \( \varepsilon \to 0 \), \( P(\varepsilon) \to \infty \) due to higher precision, and (ii) \( \varepsilon \) must ensure at least 50\% accuracy of the approximate nearest neighbor algorithm. From the worst-case approximate ratio \( \rho' = \frac{1+\varepsilon}{1-\varepsilon}\rho \), we assume \( P(\varepsilon) \leq 0 \) when \( \varepsilon \geq \frac{1}{3} \), restricting \( \varepsilon \) to \( 0 < \varepsilon < \frac{1}{3} \). Accordingly, \( P(\varepsilon) \) is defined as $P(\varepsilon) = -\ln(3\varepsilon)$. The cost function \( K(\varepsilon) \) is derived from the Johnson-Lindenstrauss lemma, which states that pairwise distances can be preserved under projection to dimension h$ \geq \mathcal{O}\left( \frac{\log M}{\varepsilon^2} \right)$. Since computational cost grows with h, we model \( K(\varepsilon) \propto \frac{1}{\varepsilon^2} \). Combining both, the benefit function becomes $B(\varepsilon) = -\ln(3\varepsilon) \cdot \varepsilon^2$, which is convex and differentiable in the feasible range. Setting \( \frac{dB}{d\varepsilon} = 0 \) yields the distortion rate $\varepsilon^* = \frac{1}{3\sqrt{e}}$.
\end{proof}

\subsection{Pseudo-code of Sample Selection}
\label{apx:pseudo_code}
Algorithms \ref{alg:query_selection} and \ref{alg:doc_selection} provide the detailed procedure of query selection and document selection, respectively.

\begin{algorithm}[t!]
\LinesNumbered  
\caption{Query Selection for Each Cluster}
\label{alg:query_selection}
\small
\DontPrintSemicolon
\KwIn{ Cluster $C_i \in \{C_1, \dots, C_k\}$ }
\KwOut{Selected query set $U$ for training}
$Q_i, D_i \gets GetClusterQueryAndDocuments(C_i)$\;
$N_i \gets GetClusterProportionalQueryCount(C_i)$\;
$U \gets \emptyset$\;
\ForEach{$q \in Q_i$}{
    $m \gets \frac{|D_i|}{|Q_i|}$
    $D_{q} \gets GetNearestDocuments(q, D, m)$\;
    $U \gets U \cup \{(q, D_q)\}$\;
}
$u \gets Random(U)$\;
\While{$|u| < N_i$}{
    $u^* \gets MaximizeCoverageQueries(u, U)$\;
    $q^*, D_q^* \gets MinimizeRedundencyQuery(u^*, u)$\;
    $u \gets u \cup \{(q^*, D_q^*)\}$\;
}
$U \gets U \cup u$\;
\Return $U$
\end{algorithm}

\begin{algorithm}[t]
\LinesNumbered  
\caption{Document Sampling for Each Query}
\label{alg:doc_selection}
\small
\DontPrintSemicolon
\KwIn{Training query  $q \in U$, cluster index $C$, number of top clusters $N$, number of negatives $k{-}1$}
\KwOut{Training document set $T_q = \{ d^+, d_1^-, \dots, d_{k-1}^- \}$}
$C_N(q) \gets getNearestClusters(q,C,N)$\;
$D_q \gets GetDocuments(C_N(q))$\;
$d^+ \gets \arg\max_{d \in D_q} Sim_{qd}$\;
$D_q^- \gets D_q \setminus \{ d^+ \}$\;
$\{ d_j^- \}_{j=1}^{k-1} \gets \arg\min^{k-1}_{d \in D_q^-} Sim_{qd}$\;
$T_q \gets \{ d^+ \} \cup \{ d_j^- \}_{j=1}^{k-1}$\;
\Return $T_q$\;
\end{algorithm}

\subsection{Time Complexity Analysis}
\label{apx:time_complexity}
We analyze the time complexity of the framework by decomposing it into four stages: cluster management, sampling, training, and retrieval. Let $Q$ denote the total number of queries up to the current session, $D$ the total number of documents, $C$ the number of clusters, $q$ the number of queries in the current session, and $p$ the number of model parameters. Cluster management involves assigning $q$ queries to $C$ clusters and decaying clusters by comparing $D$ documents to $C$ prototypes, which is dominated by $O(D)$. Sampling constructs representative queries by measuring distances between $q$ queries and $D/C$ documents, and selecting documents per query over clusters, yielding a dominant complexity of $O(D)$. Training updates the model with $p$ parameters using $q$ queries, dominated by $O(p)$, while retrieval compares $q$ queries to $D$ documents, dominated by $O(D)$. Overall, the total time complexity is $O(D + p)$.

\subsection{Details of Dataset}
\label{apx:dataset}
The dataset statistics are summarized in Table \ref{tb:dataset_summary}, which presents the domain composition (Domain), the number of queries (\#Query), the number of documents (\#Document), and the average number of relevant documents per query (\#qrels) for each dataset: LoTTE and MSMARCO. Evaluation follows a continual learning protocol over 10 sessions. Each session's training query set includes two domains: one recurring from the previous session (1) and one newly introduced (2). The 10-session structure is designed to ensure that each of the five domains appears exactly twice: once as a recurring domain and once as a newly introduced domain. The evaluation query set consists of three domains: a dropped domain not seen in the current training (i), an ongoing domain shared with the current training (ii), and a newly introduced domain (iii). For example, in the case of LoTTE, if the training query set in session $S_{t-1}$ covers Writing and Lifestyle, and the training query set in session $S_t$ covers Lifestyle (1) and Technology (2), then the evaluation query set in $S_t$ includes Writing (i), Lifestyle (ii), and Technology (iii). Depending on the evaluation setting, training and evaluation document sets are either shared (Definition~\ref{eq:staticir_evaluation}) or separated (Definition~\ref{eq:trditional_evaluation}). Domains were first distributed across the 10 sessions following this scheme, and queries were then evenly assigned. The document sets for each session were constructed to preserve the proportion of relevant documents per domain.

\begin{table}[h]
\small
\centering
\setlength{\tabcolsep}{3pt}
\caption{Datasets statistics.}
\begin{adjustbox}{max width=0.5\textwidth}
\label{tb:dataset_summary}
\begin{tabular}{l l r r r}
\toprule
\textbf{Dataset} & \textbf{Domain} & \textbf{\#Query} & \textbf{\#Document} & \textbf{\#qrels} \\
\midrule
LoTTE & Technology& 5519& 1,914,731& 6.6 \\
& Writing         & 5571& 477,066& 5.9 \\
& Lifestyle       & 5156& 388,354& 5.1 \\
& Recreation      & 5491& 430,000& 4.3 \\
& Science         & 5185& 2,037,806& 6.0 \\
\midrule
MSMARCO & Names/Public Figures & 6595& 65,860& 1.0 \\
& Dated Events         & 5960& 59,162& 1.0 \\
& Pricing/Units         & 6255& 62,517& 1.1 \\
& Medical Treatments         & 5868& 58,698& 1.1 \\
& Biology/Physics         & 6566& 65,622& 1.1 \\
\bottomrule
\end{tabular}
\end{adjustbox}
\end{table}


\subsection{Implementation Details}
\label{apx:implementation}\
We use the BM25~\cite{bm25} implementation from the Okapi library with $k_1 = 1.5$, $b = 0.75$, and $\epsilon = 0.25$. For ColBERT~\cite{colbertv2}, we use the implementation provided in the official L2R codebase. Following the original paper, we set the output dimension of the linear projection layer in the model to 128. ColBERT\textsuperscript{+} performs incremental learning using negatives sampled from BM25-retrieved (but unannotated) documents. We use the official L2R implementations of Experience Replay (ER)~\cite{er}, Maximally Interfered Retrieval (MIR)~\cite{mir}, and Gradient-based Sample Selection (GSS)~\cite{gss}. Online Coreset Selection (OCS)~\cite{ocs} is implemented based on the L2R codebase, with $\alpha = 1.0$, $\beta = 1.0$, and $\gamma = 1000.0$, following the original paper and code. For L2R~\cite{l2r}, we use the official implementation with $\alpha = 0.6$ and $\beta = 0.4$, as configured in the code.

All MCL baselines use a 30-sample memory and select one positive and six negatives per query; three negatives are sampled from the memory and three from the current batch. For L2R, we retrieve the top-50 documents with BM25 and sample from them (reduced from top-500/200 due to the smaller per-session dataset size). All DR baselines share the same encoder, \texttt{google/bert-base-uncased} (110M). Since ColBERT\textsuperscript{+} and MCL are supervised methods, we adapt them to our unsupervised setting via pseudo-labeling: for each query, we select the document with the highest cosine similarity as the pseudo-positive. For fairness, MCL replay buffers are fixed to queries from Session~0 only. All baselines are evaluated using both ground-truth and pseudo labels.

For \algname{}, we also use \texttt{google/bert-base-uncased} as the backbone. To focus on informative samples, we retain the top-50 BM25-ranked documents per query in each session. For initial cluster construction, we apply $k$-means to the first 1,024 instances, forming 12 clusters for LoTTE and 5 for MSMARCO. As defined in Equation~\ref{eq:contrastive_loss}, the similarity metric used in the loss function can be either cosine similarity or token-level similarity. Empirically, we observed no significant difference in performance between the two approaches. Therefore, we opted to use cosine similarity due to its lower computational overhead. We set the assignment factor $\lambda=8.0$ and the decaying factor $\gamma=0.25$.

\subsection{Training Time Analysis}
\begin{figure}[t]
\centering\includegraphics[width=\columnwidth]{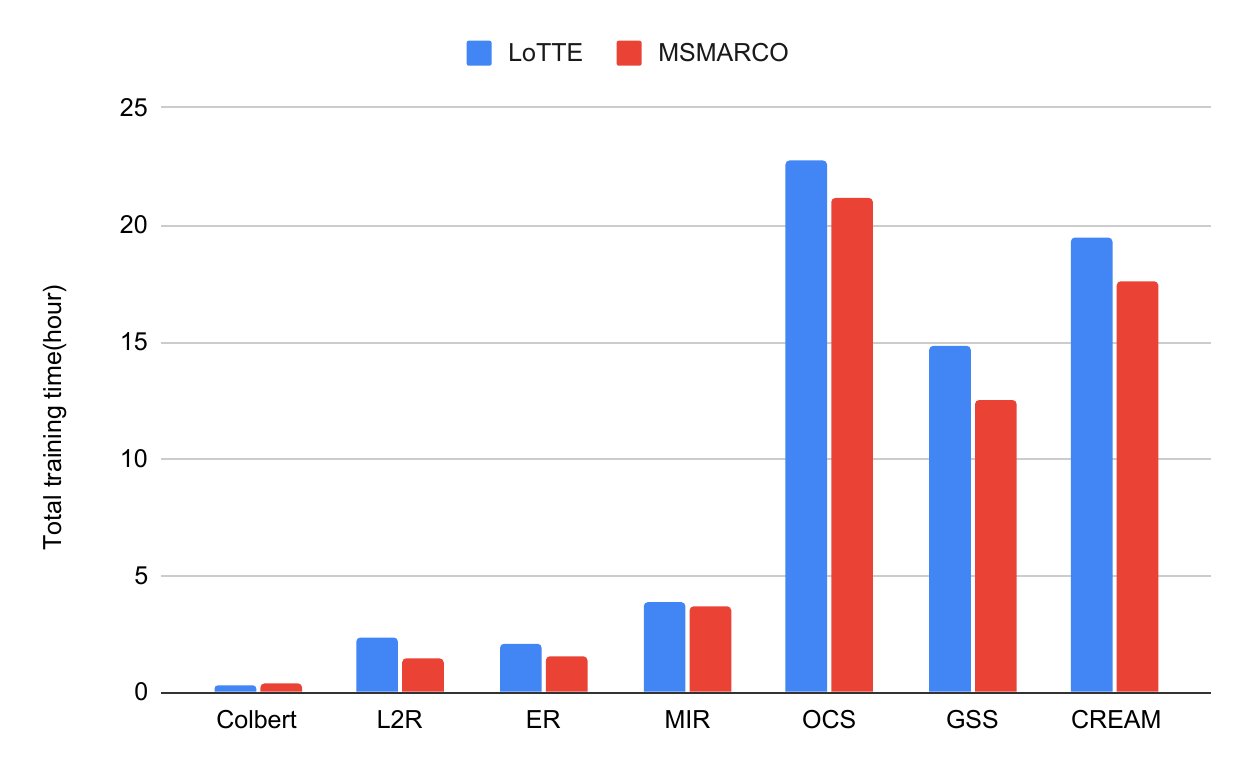}  
    \caption{Training time analysis results.}
\label{fig:training_time}
\end{figure}

As shown in Figure~\ref{fig:training_time}, 
ColBERT required the least training time, with 0.30 hours on LoTTE and 0.37 hours on MSMARCO, likely due to its use of fixed positives and negatives without any sampling strategy. Among the MCL methods, OCS incurred the highest training time—22.75 hours on LoTTE and 21.15 hours on MSMARCO—followed by GSS, which took 14.82 hours and 12.53 hours on LoTTE and MSMARCO, respectively. This can be attributed to the need to compute gradients while exploring the entire data space during sampling. In terms of overall training time, CREAM ranked second, requiring 19.48 hours on LoTTE and 17.59 hours on MSMARCO, which is also likely due to its exhaustive exploration of the data space during sampling.

\subsection{Processing Time Analysis} 
\begin{figure}[t]
\centering\includegraphics[width=\columnwidth]{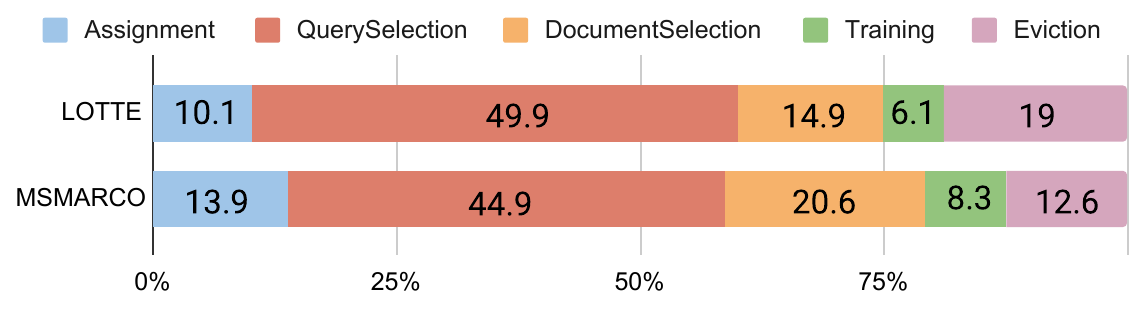}  
    \caption{Processing time analysis results.}
\label{fig:processing_time}
\end{figure}

We analyze the time consumption ratio across five processing stages: \textit{Assignment}, \textit{QuerySelection}, \textit{DocumentSelection}, \textit{Training}, and \textit{Eviction}. Figure~\ref{fig:processing_time} presents the average time and proportion spent on each stage. Among them, \textit{QuerySelection} was the most time-consuming, averaging 2.65 hours and accounting for 49\% of the total processing time, followed by \textit{DocumentSelection} (18\%) and \textit{Eviction} (14\%). The \textit{QuerySelection} and \textit{DocumentSelection} stages exhibit increasing time consumption in later sessions, as both require constructing data structures proportional to the cumulative number of queries and documents. Similarly, the \textit{Eviction} stage becomes more costly over time due to the need to identify documents to retain and re-embed the entire candidate set. All three stages (i.e., \textit{QuerySelection}, \textit{DocumentSelection}, and \textit{Eviction}) show processing times that grow with the accumulation of data across sessions. This overhead can be mitigated by tuning the parameter that controls the number of retained documents for the subsequent session.

\subsection{LSH Bit Size for MSMARCO}
\label{apx:msmarco_lsh_bit_size}
Each session includes approximately 2,430 queries (with an average of 32 tokens) and 30,000 documents (with an average of 256 tokens), resulting in up to 8 million token embeddings per session, each with 768 dimensions by BERT~\cite{bert}. Then, according to Theorem~\ref{theorem:sufficiency}, to maintain an acceptable distortion rate of $\varepsilon = \frac{1}{3\sqrt{e}} \approx 0.2$, the minimum number of RP-LSH bits required is:

\begin{equation}
\label{eq:msmarco_k_bit}
\lceil \log_2\left(\frac{8 \ln(8\times10^6)}{(0.2)^2}\right) \rceil \approx 11.
\end{equation}

\begin{figure*}[t]
    \centering
    \includegraphics[width=\textwidth]{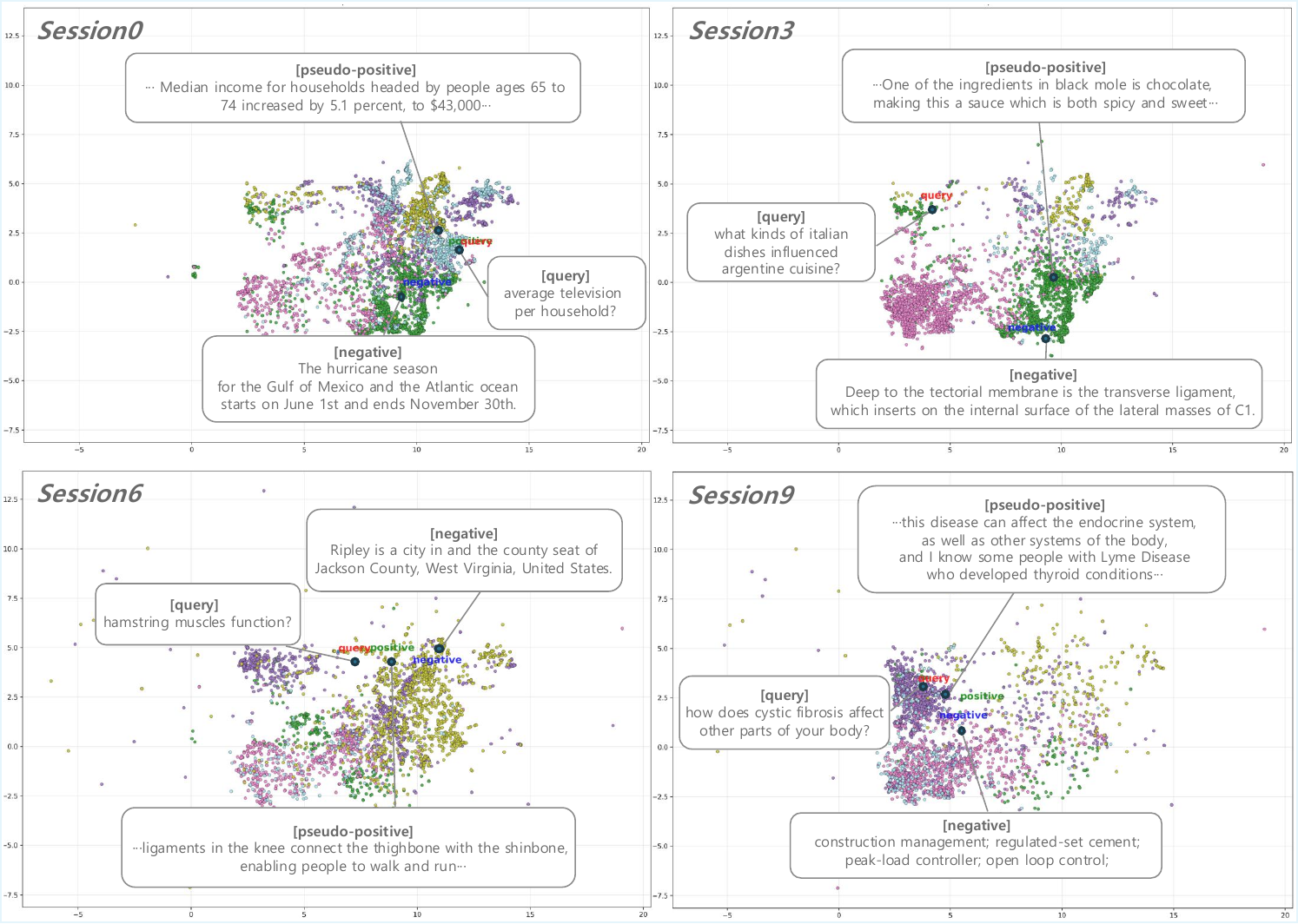}
    \caption{Soft memory with query, pseudo-positive samples, and negative samples in sessions 0, 3, 6, and 9.}
    \label{fig:visualize_memory}
\end{figure*}

\smallskip
\subsection{Qualitative Analysis of Memory Dynamics}
Figure~\ref{fig:visualize_memory} visualizes MSMARCO clusters across sessions using a UMAP projection in a shared embedding space, based on token-level similarity over 1{,}500 sampled documents per cluster. Clusters are shown with consistent colors across sessions. Queries, pseudo-positives, and negatives are marked in red, green, and blue, respectively. Pseudo-positive documents lie closer to the query than negatives, and the query–positive distance further decreases as sessions progress. These trends suggest that repeated learning on related samples helps \algname{} better capture semantic relationships, improving sentence-level matching over time. Accordingly, clusters are more intermixed early on but become more compact and better separated in later sessions, indicating increasingly well-defined topical structure.

\begin{table}[!t]
\centering
\small
\caption{Sensitivity to $\lambda$ and $\gamma$ on LoTTE and MSMARCO.}
\label{tb:additional_parameter_sensitivity}
\begin{tabular}{c|c|cc|cc}
\toprule
\multirow{2}{*}{\textbf{Parameter}}
 & \multirow{2}{*}{\textbf{Value}} & \multicolumn{2}{c|}{\textbf{LoTTE}} & \multicolumn{2}{c}{\textbf{MSMARCO}} \\
 & & S@5 & R@10 & S@5 & R@10 \\
\midrule
\multirow{2}{*}{$\lambda$}
 & 16   & 44.80 & 24.16 & 71.79 & 77.16\\
 & 4  & 47.47 & 24.60 & 70.23 & 75.86 \\
\midrule
\multirow{2}{*}{$\gamma$}
 & 0.5   & 51.40 & 27.60 & 68.79 & 74.26 \\
 & 0.125  & 42.85 & 25.81 & 64.12 & 69.09 \\
\bottomrule
\end{tabular}
\end{table}

\subsection{Sensitivity Analysis of Assignment and Decaying Factors}
\label{apx:factor_analysis}
Table~\ref{tb:additional_parameter_sensitivity} reports additional sensitivity analyses of the assignment factor $\lambda$ and the decaying factor $\gamma$ under a memory-lightweight evaluation setting with 25\% of sampling followed by BM25 top-30 filtering. Overall, the assignment factor $\lambda$ exhibited more robust performance than $\gamma$, suggesting that collecting additional documents beyond a certain threshold yields limited benefit, whereas sufficiently preserving earlier documents is critical for maintaining performance. In particular, at $\gamma = 0.125$, both LoTTE and MSMARCO suffered performance degradation, presumably because too few documents from previous sessions were retained to support learning in subsequent sessions. In contrast, increasing $\lambda$ broadens document collection, potentially capturing more useful training signal but also introducing weakly relevant noises. Thus, the assignment factor $\lambda$ reflects a trade-off between signal coverage and noise, and its optimal value may be dataset-dependent; LoTTE performed best at $\lambda=4$, while the performance on MSMARCO peaked with $\lambda=16$.